\documentclass{bjmc}   

\usepackage{cancel}
\usepackage{amssymb,latexsym,amsfonts, amsmath} 

\begin{document}


\title{The Adaptation of  Shamir's Approach for Increasing \\ the Security of a Mobile Environment
}
\titlerunning{The  security of a mobile environment}  
%
%
%
%
\author{J\=anis BULS\inst{1}, Imants GORBANS\inst{2}, Ivans KULESOVS\inst{2}, Uldis STRAUJUMS\inst{2}
}
\authorrunning{Buls, Gorbans,  Kulesovs, Straujums}   
%
%
%
%
%
%
%
\institute{Faculty of Physics and Mathematics, University of Latvia, Ze\c l\c lu iela 25, R\={\i}ga LV-1002, Latvia 
\and   Faculty of Computing, University of Latvia, Rai\c na bulv\=aris 19, R\=\i ga, LV-1586, Latvia
} 
%
%
%
%
\emails{Janis.Buls@lu.lv, Imants.Gorbans@lu.lv, Ivans.Kulesovs@gmail.com, Uldis.Straujums@lu.lv}
%
%

\maketitle      

%
\setcounter{page}{51}

\begin{abstract} 
The aim of the paper is to provide a solution which increases the
security of a mobile environment for both individuals and for 
workers in an enterprise.  The proposed solution adapts Shamir's approach for sharing a
secret for encryption key management.  One part of the key is
stored on a Bluetooth or NFC wristband or on an enterprise server, while a mobile device is
used to store all the rest.  The approach can be applied for both
securing documents and voice data. 
The solution is supported by a
mathematical formality which is missing in the currently known advice
 within cryptographic folklore.
\end{abstract}
%
%
\section{Motivation}

Despite the fact that  nobody would dispute that the storage of cryptographic keys is a very important part of data protection, there has been no significant progress in this direction as yet.
Essentially, all advice is at the cryptographic folklore level.  A digest of advice can, for example, be found in OWASP  (WEB,  \cite{Sh}):
\begin{itemize}
\item { Ensure that any secret key is protected from unauthorized access.}
\item { Define a key lifecycle.}
\item { Store unencrypted keys away from the encrypted data.}
\item {Use independent keys when multiple keys are required.}
\item {Protect keys in a key vault.}
\item {Document concrete procedures for managing keys through the lifecycle.}
\item {Build support for changing keys periodically.}
\item {Document concrete procedures to handle a key compromise.}
\item {Rekey data at least every one to three years.}
\item {Protect any keys used to secure cardholder data against disclosure and misuse.}
\item Fully document and implement all key-management processes and procedures for
cryptographic keys used for encryption of cardholder data.

The abovementioned advice mandates that key management processes should cover  8 specific key lifecycle steps:

\begin{enumerate}
\item  Generation of strong cryptographic keys.
\item  Secure cryptographic key distribution.
\item  Secure cryptographic key storage.
\item  Periodic cryptographic key changes.
\item  Retirement or replacement of keys as deemed necessary when the integrity of the
key has been weakened or keys are suspected of being compromised.
\item  Split knowledge and establishment of dual control of cryptographic keys.
\item  Prevention of unauthorized substitution of cryptographic keys.
\item  Requirement for cryptographic key custodians to sign a form stating that they
understand and accept their key-custodian responsibilities.
\end{enumerate}
\end{itemize} 

The fact that the advice on the storage of keys is at the folklore level is partially due to the lack of an adequate description of the situation which can fit into some mathematical formalism.

But, the problem is that the standard user is unable to either remember the folklore advice or to follow it. It means that a solution which is sufficiently simple has to be found, while taking into
 account all of the abovementioned restrictions. 

Our aim is to show how it's possible to ensure that both individual users and workers at an enterprise don't have to puzzle over issues which are not within their competence.
To reach this aim the authors have adopted Shamir's approach for sharing a secret (Shamir, \cite{Sha}): the secret is some data $D$. 
Divide $D$ into $n$ pieces $D_1, D_2, \ldots , D_n$
in such a way that:
\begin{itemize}
\item[(1)] knowledge of any $\varkappa$ or more $D_i$ pieces makes $D$ easily computable;
\item[(2)] knowledge of any $\varkappa-1$ or fewer $D_i$ pieces leaves $D$ completely undetermined (in the sense that all its possible values
are equally likely).
\end{itemize}
The authors propose considering the use of an encryption key as the secret data $D$ from Shamir's approach for encryption key management in a mobile environment.
We need $n=\varkappa=2$.

\section{Document Encryption} 

Split the encryption key $k$ into two parts, $k_1$ and $k_2$, so that there is no possibility of restoring a key if any of the parts are missing. 
A Bluetooth or NFC (Near Field Communication) wristband is used for storing the encrypted $k_2$. A smartphone is used for storing the remainder. That is all that the user has to care for. 
The whole document encryption process will be carried out by a mobile application on the smartphone.

Now we will describe the details that a user friendly application loaded into a smartphone must provide. 
At first a decision should be made as to which cryptosystems will be used. 

The authors propose that two different cryptosystems should be used for increasing security. 
Nowadays RC4--like stream cipher (Spritz) is considered to be good enough (see the latest research  Rivest,  Schuldt (\cite{Sp}), Ro\c sie (\cite{Ro})).   
 We do not insist on one specific cryptosystem as best of all.
For more stream ciphers see (WEB, \cite{eS}) and see (WEB, \cite{Bl}) for block ciphers. We recommend one system should be used for document encryption, the second one --- 
for communication.
Thus two different cryptosystems are loaded into the smartphone:
\begin{eqnarray*}
\mathfrak{K}_1&=&\langle \mathcal{P}_1, \mathcal{C}_1, K_1, \mathcal{E}_1, \mathcal{D}_1 \rangle,\\
\mathfrak{K}_2&=&\langle \mathcal{P}_2, \mathcal{C}_2, K_2, \mathcal{E}_2, \mathcal{D}_2 \rangle
\end{eqnarray*}
Here for both $i\in\{1,2\}$
\begin{itemize}
\item $\mathcal{P}_i$ --- plaintext space;
\item $\mathcal{C}_i$ --- ciphertext space;
\item $K_i$ --- key space;
\item  $\mathcal{E}_i:\mathcal{P}\times K\to \mathcal{C}_i $ --- a set of encryption functions; 
\item $\mathcal{D}_i:\mathcal{C}\times K\to \mathcal{P}_i $ --- a set of decryption functions
\end{itemize} 
and for every plaintext $x\in \mathcal{P}_i$, every key $k\in K_i$ is valid
 \begin{equation*} \label{f01}
\mathcal{D}_i(\mathcal{E}_i(x,k),k)=x\,.
\end{equation*}
If one cryptosystem $\mathfrak{K}_1$ is used instead of $\mathfrak{K}_1$ and $\mathfrak{K}_2$ then $K_1\not= K_2$ at least .

For every document $D_i$ a cryptosystem $\mathfrak{K}_1$ key $K_{i1}$ is generated, which is used for document $D_i$ encryption, namely,
\[
D'_i=\mathcal{E}_1(D_i,K_{i1}).
\]
Key $K_{i1}$ is split into two independent parts $K_{i1}'$  and $K_{i1}''$ so that 
\[
K_{i1}=\mathfrak{A}(K_{i1}',K_{i1}''),
\]
where $\mathfrak{A}$ --- algorithm, restoring  key $K_{i1}$, if its parts $K_{i1}'$ and $K_{i1}''$ are given.
The requirement is that key $K_{i1}$ cannot be restored, if only one part of the key parts  $K_{i1}'$, or $K_{i1}''$ is known.

 Cryptosystem $\mathfrak{K}_2$ keys $K_{i2}'$ and $K_{i2}''$ are generated. The halves of  key $K_{i1}$ are encrypted, namely,
\begin{eqnarray*}
S_{i1}&=&\mathcal{E}_2(K_{i1}',K_{i2}'),\\
S_{i2}&=&\mathcal{E}_2(K_{i1}'',K_{i2}'').
\end{eqnarray*}

$D'_i,  K_{i2}'', S_{i1}$ are  stored in the smartphone, while $ K'_{i2}, S_{i2}$ are stored in the Bluetooth or NFC wristband (see Fig.1.). 

\special{em:linewidth 0.4pt}
\unitlength 1.00mm
\linethickness{0.4pt}
\begin{picture}(130,80)(15,0)

\put(50,75){\line(1,0){50}}
\put(50,75){\line(0,-1){20}}
\put(100,75){\line(0,-1){20}}
\put(50,55){\line(1,0){50}}

\put(65,69){\makebox(0,0)[lc]{Mobile device}}
\put(67,61){\makebox(0,0)[lc]{$D'_i,  K_{i2}'', S_{i1}$}}

\put(50,35){\line(1,0){50}}
\put(50,35){\line(0,-1){20}}
\put(100,35){\line(0,-1){20}}
\put(50,15){\line(1,0){50}}

\put(56,30){\makebox(0,0)[lc]{Programmable Bluetooth}}
\put(64,26){\makebox(0,0)[lc]{or NFC
device}}

\put(68,20){\makebox(0,0)[lc]{$ K'_{i2}, S_{i2}$}}

\put(73,35){\line(0,1){20}}  \put(73.02,50){\vector(0,1){0.2}}

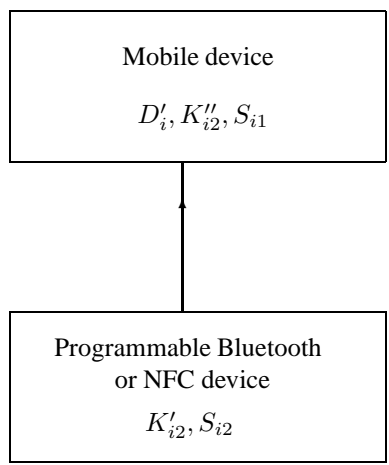
\put(58,5){\makebox(0,0)[lc]{Fig.1. Splitting of the key.}}
\end{picture}

If a user wants to read the document $\mathcal{D}_i$, then the user has to enter the password 
(a security measure for the situation where a  mobile device is also used by other persons, let's say, family members), and only then does the application address the wristband,
 loading $K'_{i2}, S_{i2}$,
and decrypting the halves of key $K_{i1}$
\begin{eqnarray*}
K'_{i1}&=& \mathcal{D}_2(S_{i1}, K'_{i2}),\\
K'_{i2}&=& \mathcal{D}_2(S_{i2}, K''_{i2}),
\end{eqnarray*}
restoring the document's $D_i$ encrypted key 
\[
K_{i1}=\mathfrak{A}(K'_{i1},K'_{i2});
\]
and finally decrypting the document 
\[
D_i=\mathcal{D}_1(D'_i, K_{i1}).
\]

As soon as the document is decrypted, the application deletes (destroys) all the data used for the decryption of the document  which should not be stored on the mobile device, namely, 
the keys 
$K_{i1}, K'_{i1},K''_{i1}, K'_{i2},K''_{i2}$
 and  encrypted keys $S_{i1},S_{i2}$. 
Work with memory must be organized depending of the specific hardware to really destroy mission critical data.
Decrypted document  $D_i$ is deleted (destroyed) when the user finishes reading 
the document. Only $D'_i,  K_{i2}'', S_{i1}$ should remain on the smartphone 
after the particular document has been read  (see Fig. 2.). Otherwise after reading the document becomes unsecure.

\special{em:linewidth 0.4pt}
\unitlength 1.00mm
\linethickness{0.4pt}
\begin{picture}(130,80)(15,0)

\put(5,0){
\put(5,75){\line(1,0){40}}
\put(5,75){\line(0,-1){20}}
\put(45,75){\line(0,-1){20}}
\put(5,55){\line(1,0){40}}
\put(5.7,70){\makebox(0,0)[lc]{Programmable Bluetooth}}
\put(14,66){\makebox(0,0)[lc]{or NFC device}}
\put(18,60){\makebox(0,0)[lc]{$ K'_{i2}, S_{i2}$}}
}

\put(55,70){\makebox(0,0)[lc]{$ K'_{i2}$}}
\put(50,65){\line(1,0){15}}  \put(60.02,65){\vector(1,0){0.2}}
\put(55,60){\makebox(0,0)[lc]{$  S_{i2}$}}

\put(71,72){\makebox(0,0)[lc]{Mobile device}}
\put(74,66){\makebox(0,0)[lc]{$D'_i,  K_{i2}'', S_{i1}$}}
\put(76,60){\makebox(0,0)[lc]{$ K'_{i2}, S_{i2}$}}

\put(65,75){\line(1,0){35}}
\put(65,75){\line(0,-1){20}}
\put(100,75){\line(0,-1){20}}
\put(65,55){\line(1,0){35}}

\put(110,70){\makebox(0,0)[lc]{key restoring}}
\put(100,65){\line(1,0){33}}  \put(120,65){\vector(1,0){0.2}}
\put(110,60){\makebox(0,0)[lc]{decrypting}}

\put(4,0){
\put(6,30){\line(1,0){12}}  \put(13,30){\vector(1,0){0.2}}
\put(18,45){\line(1,0){29}}
\put(18,45){\line(0,-1){30}}
\put(47,45){\line(0,-1){30}}
\put(18,15){\line(1,0){29}}

\put(21,42){\makebox(0,0)[lc]{Mobile device}}
\put(24,34){\makebox(0,0)[lc]{$D'_i,  K_{i2}'', S_{i1}$}}
\put(26,28){\makebox(0,0)[lc]{$ K'_{i2}, S_{i2}$}}
\put(20,22){\makebox(0,0)[lc]{$ K'_{i1}, K''_{i1}, K_{i1}, D_i$}}

\put(49,33){\makebox(0,0)[lc]{reading}}
}

\put(47,0){
\put(4,30){\line(1,0){14}}  \put(13,30){\vector(1,0){0.2}}
\put(18,45){\line(1,0){29}}
\put(18,45){\line(0,-1){30}}
\put(47,45){\line(0,-1){30}}
\put(18,15){\line(1,0){29}}
}

\put(0,0){
\put(68,42){\makebox(0,0)[lc]{Mobile device}}
\put(71,34){\makebox(0,0)[lc]{$D'_i,  K_{i2}'', S_{i1}$}}
\put(73,28){\makebox(0,0)[lc]{$ \cancel{K'_{i2}}, \cancel{S_{i2}}$}}
\put(67,22){\makebox(0,0)[lc]{$ \cancel{K'_{i1}}, \cancel{K''_{i1}}, \cancel{K_{i1}}, D_i$}}
}

\put(94,30){\line(1,0){15}}  \put(104,30){\vector(1,0){0.2}}
\put(90,0){
\put(19,45){\line(1,0){24}}
\put(19,45){\line(0,-1){30}}
\put(43,45){\line(0,-1){30}}
\put(19,15){\line(1,0){24}}
}

\put(40,0){
\put(71,42){\makebox(0,0)[lc]{Mobile device}}
\put(74,34){\makebox(0,0)[lc]{$D'_i,  K_{i2}'', S_{i1}$}}
\put(82,22){\makebox(0,0)[lc]{$ \cancel{D_i}$}}
}

\put(97,33){\makebox(0,0)[lc]{after}}
\put(97,27){\makebox(0,0)[lc]{reading}}

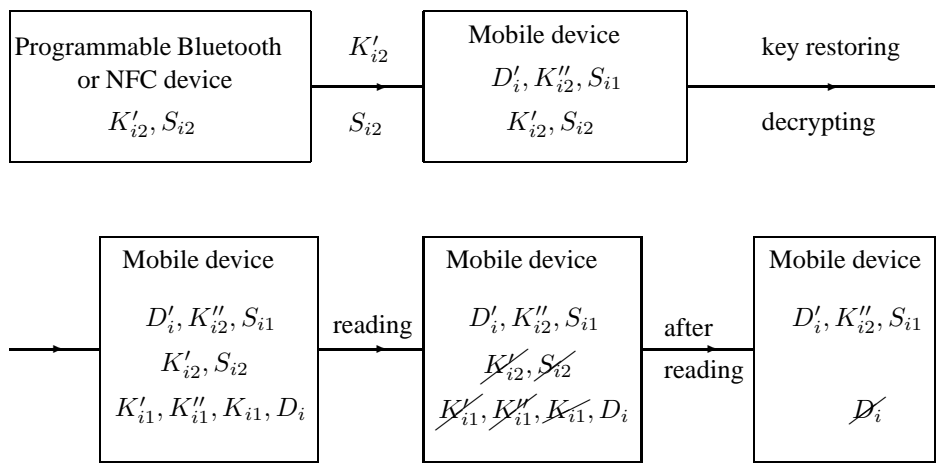
\put(58,5){\makebox(0,0)[lc]{Fig.2. Reading of document $D_i$.}}
\end{picture}

Where a mobile device has been lost, unauthorized access to encrypted documents is not possible, assuming that the wristband has also not been lost.
Our solution does not exclude the usage of other security measures, see, for example (WEB, \cite{Gar});  Gorbans,  Kulesovs,  Straujums,  Buls (\cite{Gor}). 

\section{Encryption of Enterprise Documents and Voice Calls}

Users can use their connection with an enterprise server instead of the wristband for the encryption of enterprise documents.  This eliminates the risk arising from the loss of a wristband
and allows the disabling of communication with the lost or stolen mobile device. In this case, the keys are generated on a secure enterprise server. 
$K'_{i2}$ and $S_{i2}$ are kept there now as well. Using this approach, a user has to take into account that encrypted documents are not available offline, but this issue should be 
solved separately.


As opposed to the reading of documents, voice calls should be encrypted and decrypted in real time to avoid delays.
In general, there are no serious barriers to encrypting voice calls nowadays. It should be noted, that mobile networking already goes from an analogue 
to a digital mode, and traffic is encrypted if the call moves within an LTE network. But, the encryption process is controlled by the carrier's mobile tower. 
That is why an enterprise cannot be 100\%  sure that encryption has really taken place between the phone and the tower, and that the call has also been transferred in 
an encrypted way between other towers, especially if they belong to another carrier.

In order to provide enterprise users with operative networking, we recommend a solution similar to the one already proposed for private needs. Two different cryptosystems 
are uploaded to the server:
\begin{eqnarray*}
\mathfrak{K}_3&=&\langle \mathcal{P}_3, \mathcal{C}_3, K_3, \mathcal{E}_3, \mathcal{D}_3 \rangle,\\
\mathfrak{K}_4&=&\langle \mathcal{P}_4, \mathcal{C}_4, K_4, \mathcal{E}_4, \mathcal{D}_4 \rangle.
\end{eqnarray*}

Let's assume that communication should be established between $\nu$ employees. 
The $m$ sets are prepared for each pair of $i,j$ employees. The first set is used for the first call, the second set --- for the second call, etc. If $i$ employee calls $j$
employee for time  $k$, then the application  automatically loads the 
$k$-th set $ K^{k'}_{ij4}, S^k_{ij2}$ from the wristband and renews
 the key $K^k_{ij1}$. This key is used for call encryption and decryption. When the call is completed or if the connection has been unsuccessful, then
 set $K_{ij4}^{k'}, K_{ij4}^{k''}, S_{ij1}^k,S_{ij2}^k$
 is deleted.

 The $\frac{m\nu(\nu-1)}{2}$ various key sets are generated $K_{ij3}^t,  K^{t'}_{ij4}, K^{t''}_{ij4}$, 
where $K_{ij3}^t$ is the cryptosystem
$\mathfrak{K}_3$ key used for the encryption of a call $t$ between $i$ and $j$ employees. Keys $K^t_{ij3}$ are divided into two independent parts $K^{t'}_{ij3}$ and $K^{t''}_{ij3}$.
 It is done in such a way that  
\[
K^t_{ij3}=\mathfrak{A}(K^{t'}_{ij3},K^{t''}_{ij3}),
\]
 where $\mathfrak{A}$ is an algorithm that renews key $K^t_{ij3}$ if its parts, $K^{t'}_{ij3}$ and $K^{t''}_{ij3}$, are given.

Keys $K_{ij4}^{t'}$ and $K_{ij4}^{t''}$ of the cryptosystem $\mathfrak{K}_4$ are generated. The halves of key $K_{ij3}^t$ are encrypted, namely,
\begin{eqnarray*}
S_{ij1}^t&=&\mathcal{E}_4(K_{ij3}^{t'},K_{ij4}^{t'}),\\
S_{ij2}^t&=&\mathcal{E}_4(K_{ij3}^{t''},K_{ij4}^{t''}).
\end{eqnarray*}
$ K_{ij4}^{t''}$ and  $S^t_{ij1}$ are kept on the smartphone while $ K^{t'}_{ij4}$ and  $S^t_{ij2}$ are kept on the wristband.

There is one encryption characteristic worth mentioning. If there is no need to keep a recorded call in an encrypted way, then such a protocol is not needed either. In this case it is enough
 to choose one cryptosystem $\mathfrak{K}_3$ and to prepare $m$ encryption keys  $K^t_{ij3}$ for each $i,j$ pair. It is clear that if  call $k$ has taken place or a connection has been 
unsuccessful, then key $K^k_{ij3}$ must be destroyed.

The enterprise server serves the main role in call encryption key generation and the initial distribution. Furthermore, the distribution of keys occurs at the premises of the enterprise. 
If the number 
of employees $\nu$ is not very large (not exceeding, for example, 100) and an attack that blocks calls is not expected, then the protocol offered is practically secure.

A more severe risk is connected to the key generation algorithm. The following situation could be an example of risk exposure. Let's suppose that the procedure generates only 
$2^{40}$ keys, when it was advertised to users that the generator would generate $2^{256}$ independent keys. The user cannot identify the fraud in a direct way, because it could take 
quite a lot of time to test the generator. Just imagine if, for example, each key is generated in one second. This is why we recommend that enterprises should create key generators themselves.

If there is no generator software available, then key generation can be done
manually.
 Anyway each cryptosystem need keys with specific
properties. We shall demonstrate the method how can be generated a key that contains approximately equal number 0 and 1 in the key.
 A key could be generated using a card deck. Take 52 cards, mix them up, and
throw one or two cards out. Take the cards from the deck one by one. If a card is red, then choose 0.
If a card is black, then choose 1.
A key (or part of key) with a length of 50 or 51 is generated. Repeat the whole procedure just described again and combine the parts. A key with a length of 203 
(4 parts combined) is considered to be practically secure.

 
The method suggested could be seen to be comically primitive, and thus not secure by a non-specialist. But, 
\[
\left(\begin{array}{c}
2n\\ n
\end{array}\right) \sim\frac{4^n}{\sqrt{\pi n}}
\]
This is well known  asymptotic approximation (see, for example  (WEB,  \cite{Wik})).   
For standard 52--card deck $n=\frac{52}{2}=26$.
We recommend to throw one or two cards out  therefore  it is possible to generate about 
\[
\frac{(4^{25})^4}{(\pi 25)^2}\ge 2^{187}
\]
different keys using the suggested method.  Such a method could be used by a user who needs to encrypt 2--3 documents. 
But this method is not suitable for an enterprise which needs to encrypt  thousands of documents each day.

\section{Conclusions}

We have adapted Shamir's approach to share a secret (Shamir (\cite{Sha})) for encryption key management in a mobile environment.
An encryption key of sufficient length is generated and split into two independent parts. This is used to encrypt a document or voice call. Part of the key is kept in a smartphone, 
while another part is kept in a Bluetooth or NFC wristband. Part of the key is kept in the wristband 
which is why decryption of a document is not possible without that part.

The fact that part of the key is kept in a wristband in an encrypted way prevents an attacker from gaining  any significant benefit, even if traffic between the wristband and 
smartphone is intercepted. That is why a user should not be worried if someone illegally intercepts that traffic. Taking Bluetooth characteristics into consideration,
the illegal interception of traffic is possible from  quite a long distance. NFC technology is less vulnerable to this risk.

A secure server could be used instead of a wristband for the needs of enterprises.

\section{Acknowledgements}

This research is part of a project called "Competence Centre of Information and Communication Technologies" run by the IT Competence Centre, 
Contract No.L-KC-11-0003, Activity No.1.22, co-financed by European Regional Development Fund.

%

%
%

%
%

\end{document}